\documentclass[runningheads, 12pt]{llncs}
\usepackage{times}
\usepackage[left=3cm, right=3cm, top=3cm, bottom=3cm]{geometry}
\usepackage{rotating}
\usepackage{subcaption} 
\usepackage{booktabs} 
\usepackage{graphicx}
\usepackage{subcaption}
\usepackage{amssymb}
\usepackage{amsmath}
\usepackage{xcolor}

\usepackage{tikz}
\usetikzlibrary{arrows,calc,positioning}

\begin{document}
\title{Smart Data-Driven GRU Predictor for\\
SnO2 Thin films Characteristics\thanks{LPCMIA, LITIO, LINFI, and FEMTO-ST, *Corresponding Author.} \\
}

\author{Faiza Bouamra\inst{1} \and
Mohamed Sayah\inst{2,*} \and
Labib Sadek Terrissa\inst{3} \and Noureddine Zerhouni\inst{4}}
\authorrunning{Faiza Bouamra, Mohamed Sayah, and al.}
\institute{LPCMIA Laboratory, University of Blida1, Sciences Faculty, Physics Department, BP270 route de Soumaa, Algeria \and
LITIO laboratory, University of Oran1, Department of Computer Science, BP1524, El Mnaouer, Es-senia, Oran, Algeria
\and
LINFI Laboratory, Department of Computer Science, University of Biskra, BP145 RP 07000, Algeria
\and 
FEMTO-ST, Franche-Comté Electronique Mécanique Thermique et Optique – ENSMM, France\\
\and
\textbf{Publication Baltica Journal, Special Issue, September 2024, p263-281, Extended version ICCAP2023.}
}
\maketitle              
%
\begin{abstract}
In material physics, characterization techniques are foremost crucial for obtaining the materials data regarding the physical properties as well as structural, electronics, magnetic, optic, dielectric, and spectroscopic characteristics. However, for many materials, ensuring availability and safe accessibility is not always easy and fully warranted. Moreover, the use of modeling and simulation techniques need a lot of theoretical knowledge, in addition of being associated to costly computation time and a great complexity deal. Thus, analyzing materials with different techniques for multiple samples simultaneously, still be very challenging for engineers and researchers. It is worth noting that although of being very risky, X-ray diffraction is the well known and widely used characterization technique which gathers data from structural properties of crystalline 1d, 2d or 3d materials.
We propose in this paper, a Smart GRU for Gated Recurrent Unit model to forcast structural characteristics or properties of thin films of tin oxide SnO2(110). Indeed, thin films samples are elaborated and managed experimentally and the collected data dictionary is then used to generate an AI - Artificial Intelligence - GRU model for the thin films of tin oxide SnO2(110) structural property characterization. 

\end{abstract}


\section{Introduction}
\label{intro}
X-Ray diffraction, discovered by Röntgen in 1895~\cite{mittemeijer2010fundamentals}, is a crucial method for understanding and extracting structural properties of crystalline materials, which are usually resumed in lattice parameters a, b, c, grain size D and strains~\cite{sivakumar2021effect}. This diffraction technique plays a pivotal role in characterizing material properties across various fields, from minerals to advanced technological raw materials. The X-Rays diffraction principle consists on the interaction of X-rays with solid materials, resulting in diffraction patterns that provide insights into the periodicity of atomic structures. Unlike this, availability and exponential cost of X-ray diffraction make its use very challenging for researchers and engineers, but mainly because that X-ray diffraction requires significant samples of experimental data due to the high number of parameters. Furthermore, the \texttt{ab-initio} calculation methods  are emerging nowadays as a promising alternative for characterizing the atomic structure and properties of materials.

Unfortunately and instead of the progress of theoretical and simulation methods, the structural characterization still be too limited because of the current available computing capacity of machines, especially when dealing with complex crystalline structures with high number of atoms~\cite{saxena2008damage12}~\cite{ali2022x}. Thus, more new explorations have been done to search for soft and less costly alternatives, particularly around the fields of advanced simulation and artificial intelligence techniques that were enhanced by the evolution in modeling approaches but also technological computing capacities. In fact, new and innovative data-driven methods were considered to estimate or even forecast physical properties of materials. These well known artificial intelligence (AI) techniques are now widely adopted in various domains such as machinery, chemistry, physics, and PHM (Prognostics and Health Management)~\cite{si2011remaining}\cite{mosallam2015component}\cite{si2017data4}\cite{heimes2008recurrent}\cite{tian2012artificial}. 

We notice that when studying materials and dealing with complex crystalline structures in the context of limited access to equipment, artificial intelligence networks such as the GRU - Gated Recurrent Unit models and others are gaining prominence, value and trust and are currently considered as the more favorite techniques to enhance XRD's capabilities and strengthen predictions of materials' structural properties~\cite{jesche2016x}. Indeed, it is worth noting that in our context, thin films samples are elaborated experimentally by the technique Sol-Gel Dip-Coating \cite{tang2017dip}~\cite{brinker2013dip} with some defined conditions such as catalysis, velocity of dip, time of dip, number of dip. The obtained samples are characterized with X-ray diffraction technique and then analyzed with HighScore tool~\cite{degen2014highscore}, to have the structural parameters a, b and D. Next, using the collected data samples dictionary from the real experiences, our proposed GRU neural network Predictor of the structural properties, is generated after a learning and testing processes. Finally, the designed AI-based GRU model~\cite{shiri2023comprehensive} along with the elaborated experiments show the benefits of our approach for material characterization and prove the relevance and insights of implementing data-driven models for material properties prediction. Later in the recent future, the use of more robust and complex AI models should certainly boost the material characterization domain and this will cover a wide range of materials. 

\section{Data-Driven Material Characterization Framework}
\begin{figure}[h]
    \centering
    \includegraphics[scale=0.4]{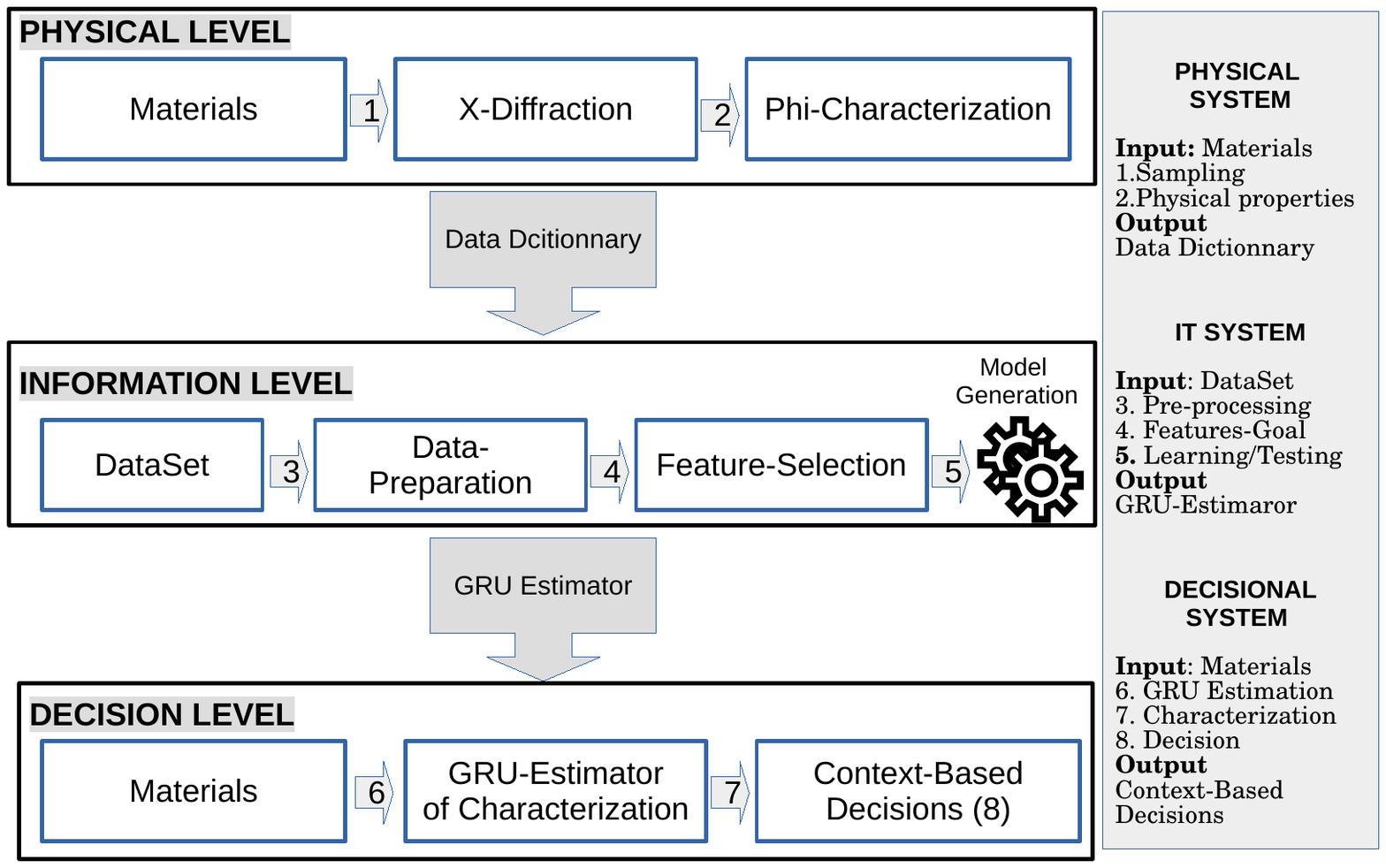}
    \caption{AI-Based Materials Characterization Framework.}
    \label{fig:Characterizingframe}
\end{figure}

In the depicted framework below (see~Figure~\ref{fig:Characterizingframe}), we illustrate the different stages of our data-driven characterization process. When fully developed, the proposed AI GRU model~\cite{sayah2020GRUThinFilmsSnO2} will facilitate the materials characterization by estimating the cristal properties using the experimental dataset.

We notice that there are three different stages or modules interrelated in order to help experts in their final decisions:

\begin{enumerate}
    \item \textit{Experiment Stage} : At this stage - see Figure~\ref{fig:Characterizingframe}, the engineers and researchers do real physical experiments, define relevant properties that may fit the physical or chemical situation and collect concrete samples to construct a data dictionary that summarises the experimental specifications. These collected data are then used in modeling, simulation, or even to design data-driven models.  
    
    \item \textit{Information Stage} : This stage - Figure~\ref{fig:Characterizingframe} - mainly concerns the data-driven model development. The data dictionary is pre-processed and the data are prepared using normalization, regularization and features reduction to respond to a task or goal oriented request. Finally an AI  model is designed and tested to get better of its performances.
    
    \item \textit{Decision Stage} : In this stage we integrate the constructed AI driven model in the decision process. Here, the data-driven model will be tested to see its compliance by regards to the client and user requirements.   
\end{enumerate}

\section{X-ray Diffraction for SnO2 Characterization}
\label{X-rayDiff}

\begin{figure}[h!]
    \centering
    \includegraphics[scale=2.2]{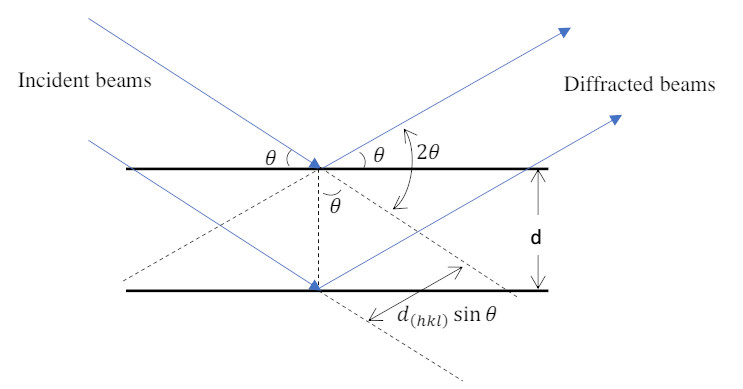}
    \caption{XRD Diffraction Scheme.}
    \label{fig:diffracte}
\end{figure}

In this section, we will describe in details the process of structure materials characterization using the non destructive X-ray diffraction technique~-~see Figure~\ref{fig:diffracte}. This latter shows a derivation of Bragg's law, where a sample of crystal material is placed at the center of the instrument, and an incident beam of X-ray is directly oriented towards this sample-~see details in Figure~\ref{fig:diffracte}. Then, the outgoing beam, who makes a $2\theta$ with awards one, is obtained as a result of interaction with the atoms of the crystal. Every wavelength $\lambda$ of the X-ray beam should be proportional to the spacing of the set of crystal planes ($hkl$), d and the reflection will occur if Bragg’s law below is satisfied - See the equation~\ref{eq:bragg}, where, $n$ is a positive integer, $\lambda$ a wavelength and $d$ is gain size.  

\begin{equation}
    2dsin(\theta) = n\lambda,~~~\\
    Where~~~\lambda = 1.540 A^{\circ}
    \label{eq:bragg}
\end{equation}

Next, to compute lattice parameters a and c, we use the equation~\ref{eq:tertaXrystal1}~obtained from lattice spacing formula for a tetragonal crystal system \cite{cullity1956elements}, where h,k and l are Miller indices defining an (hkl) plane in the lattice : 

\begin{equation}
    \frac{1}{d^2} = \frac{h^2 + k^2}{a^2} + \frac{l^2}{c^2}
    \label{eq:tertaXrystal1}
\end{equation}

By applying the Bragg law’s and considering the constants A = $\lambda^2/(4a^2$)  and C = $\lambda^2/(4c^2$), we obtain:

\begin{equation}
    sin^2(\theta) = A(h^2 + k^2) + Cl^2.\\
\end{equation}

Thus, for $l = 0$, we define $hk0$ as $sin^2(\theta) = A(h^2 + k^2)$, and from the other lines of the patterns, we determine the C value in : $sin^2(\theta) - A (h^2 + k^2) = C l^2$. Finally, the Grain size D, is calculated using Scherrer relation~\cite{drits1997xrd}: 

\begin{equation}
    D = \frac{K*\lambda}{\beta*cos(\theta)}
    \label{eq:tertaXrystal}
\end{equation}

With K the Scherrer constant, on which depends the crystallites, and is typically in [0.9, 1.0]. $\beta$ is the full width at the half-maximum (FWHM).\\

\begin{figure}[h]
    \centering
    \includegraphics[scale=0.45]{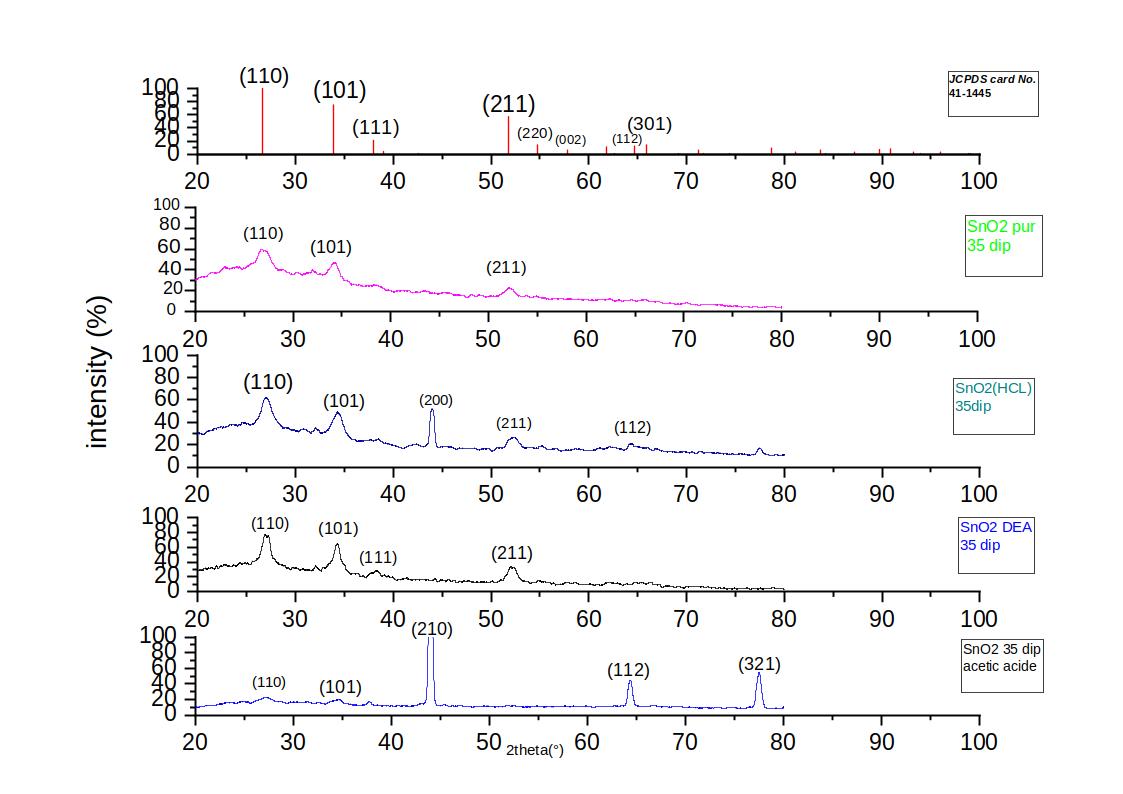}
    \caption{X-Ray diffraction patterns of SnO2 thin films obtained by Sol-Gel (Dip Coating) technique. a) ideal pattern of Sno2, b) pur SnO2, c) SnO2 with HCI solution, d) SnO2 with DEA solution, e) SnO2 with acetic acide.}
    \label{fig:spectra}
\end{figure}

We notice that the obtained patterns has interference maxima (peaks) corresponding to the diffracted X-ray by the regularly planes of the crystal lattice - see Figure~\ref{fig:spectra}. It's worth nothing to know that X-ray diffraction techniques involves two main methods resumed in (1) powder or (2) single crystal diffraction. \\

In our research work, we especially have a focus on the diffraction of X-ray for single crystal of thin oxide SnO2. In addition, samples of SnO2 thin films were elaborated with sol-gel dip coating technique. After getting a reliable thin films, we have optimized some experimental parameters such as solution, speed of dip, number of dip and others and finally we get these parameters that correspond to the final output sample with its proper characteristics.

In the Figure~\ref{fig:spectra}, we represent some results patterns from X-ray diffraction technique that were treated with HighScore software \cite{degen2014highscore}. The first pattern represent theoretical sheet obtained from data based on the perfect SnO2 thin films, and the others are the different cases obtained with changing experimental parameters, as mentioned clearly in the Figure~\ref{fig:spectra}. These collected data are considered and examined during the development of the GRU model.

\section{GRU-Based Predictor}
\label{methodos}
\begin{figure}[h!]
    \centering
    \includegraphics[scale=.95]{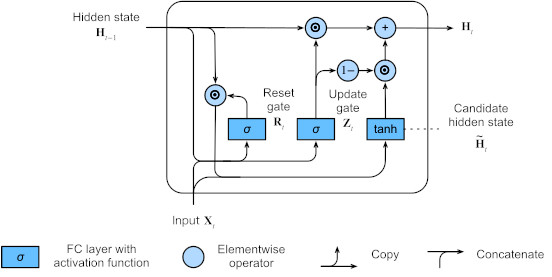}
    \caption{GRU Cell/Unit Computation\cite{xu2020deep}.}
    \label{fig:2}
\end{figure}

As a streamlined version of complex AI models~\cite{sherstinsky2020fundamentals}\cite{sayah2021robustness}\cite{sayah2021deep}\cite{remadna2022boosting}\cite{remadna2022rul}~, the gated recurrent unit (GRU)~\cite{cho-etal-2014-properties} achieves good performance and do faster computing. A GRU model contains a reset gate and an update gate. These gates are composed by sigmoid activations functions $\sigma$ to force their output values to lie in the interval [0, 1]. The reset gate controls at what extent the previous states should we remembered. In the same way, the update gate tracks how much of the new states are just a copy of the old ones~- See Figure~\ref{fig:2}. 

Here are the analytical equations that map the GRU cell’s computation states at each time step $t$. Let consider the notations, where W$_*$, W'$_*$, and b$_*$ denote respectively the learned weight matrices and biases for each gate referred to as index $*$. Also, the $\sigma$ is the logistic sigmoid function. The current input is $X_t$, and R, Z  are the reset and update gates respectively, $H_{t-1}$ denotes the previous iteration's hidden state, whereas $H_t$ corresponds to the current hidden state. 

\begin{equation}
\begin{split}
Z_t &= \sigma(W_Z \: X_t + W'_Z \: H_{t-1} + b_Z) \\
R_t &= \sigma(W_R \: X_t + W'_R \: H_{t-1} + b_R) \\
G_t &= \tanh(W_G \: X_t + W'_G \: (R_t \circledcirc H_{t-1}) + b_G) \\
H_t &= Z_t \circledcirc H_{t-1} + (1-Z_t) \circledcirc G_t
\end{split}
\label{eq:yourlabel}
\end{equation}

The flowchart below illustrates the inputs for both the reset and update gates in a GRU, given the input of the current time step $X_t$ and the hidden state $H_{t-1}$ of the previous time step. The outputs of the gates are given by two fully connected layers with a sigmoid activation function.

\subsection{Performance Measurements}
The adopted performance metrics below will be used to depict and to prove the performance of the proposed regression GRU model, and this enables the identification of a good performing parameters that best fits the requirements of our defined problem. In fact, the mean squared error $MSE$ in the equation~\ref{eq:MSE} computes the error between the experimental $c$ ground truth values and the GRU estimated  $c'$ values. 

\begin{equation}
   MSE=\frac{1}{n}\Sigma_{i=1}^{n}{{\Big({c_i'-c_i}\Big)^2}}
   \label{eq:MSE}
\end{equation}

Again, the $R^2$ score, the Coefficient of determination, is a regression metric. This metric is a normalized version of the MSE that makes it easier to compare ML models. The closer $R^2$ is to 1 (ideal), it means the regression was able to capture 100
\begin{equation}
  R^2 = 1 - \frac{\Sigma_{i=1}^{n}{\Big({c_i'-c_i}\Big)^2}}{\Sigma_{i=1}^{n}{\Big({c_i-\overline c}\Big)^2}} 
  \label{eq:R2}
\end{equation} 

With $n$ the total number of samples in testing data set, and $c$, $c'$ respectively the actual coefficient value and  the estimated coefficient. \\

In addition, the scores $S$ measurement in ~\ref{eq:Score} formula track the sensitivity of the learning and shows its smoothness.

\begin{equation}
\centering
    S = \left\{\begin{split}
         &\alpha*(\frac{exp^{|ei|}}{10} -1) & |e_i| >= \epsilon \\
         &0 & |e_i| < \epsilon 
        \end{split}\right.
       \label{eq:Score}
 \end{equation}
 
Where, $S$ is the sensitivity score of the learning/generalization functions. $\alpha$ is the amplification factor, and $e_i$ is the deviation between the learning \texttt{loss} and the generalization \texttt{val-loss} parameters. When, the absolute positive error $|e_i|$ is exactly less than $\epsilon$ the sensitivity tends to zero ($0$) and the learning/generalization process is good. Conversely, In contrast, for every error $|e_i|$ greater or equal to $\epsilon$, the learning is not stable even if the \texttt{loss} decreases considerably.  

\section{GRU Model: Design and Experiments}
\label{GRUModel}

In this section, we design a GRU-Based Predictor for structural properties of SnO2, and its possible adoption for the characterisation of various other materials. Clearly, we depicted in the following the different steps that are used to construct, tune and test the GRU model, starting by the essential data collection and preparation phase for each AI data-driven modeling.

\tikzstyle{intg}=[draw,minimum size=3em,text centered,text width=6.cm]

\begin{figure}[ht]
    \centering
    \begin{tikzpicture}[
      >=latex',
      auto
    ]
      \node [intg] (Step1)  {\textbf{Step 1}: Problem Identification $\mathcal{P}$ ?};
      \node [intg]  (Step2) [node distance=1.5cm, below of=Step1] {\textbf{Step 2}: Collect Data samples $\mathcal{D}$ that describe the Context of the $\mathcal{P}$ Problem};
      \node [intg]  (Step3) [node distance=1.5cm ,below of=Step2] {\textbf{Step 3}: Data Pre-Processing of $\mathcal{D}$};
      \node [intg] (Step4) [node distance=1.5cm,below of=Step3] {\textbf{Step 4}: Feature Selection on $\mathcal{D}$ for a given goal $g$};
      \node [intg] (Step5) [node distance=1.5cm,below of=Step4] {\textbf{Step 5}: Design and Tune Network AI model using $<\mathcal{P}$, $\mathcal{D}>$ and get best performance that better fits the goal $g$ };

      \draw[->] (Step1) -- ($(Step1.south)$) -| (Step2) node[above,pos=0.5] {};
      \draw[->] (Step2) -- ($(Step2.south)$) -| (Step3) node[above,pos=0.5] {};
      \draw[->] (Step3) -- ($(Step3.south)$) -| (Step4) node[above,pos=0.5] {};
      \draw[->] (Step4) -- ($(Step4.south)$) -| (Step5) node[above,pos=0.5] {};
      
    \end{tikzpicture}
    \caption{GRU-Based Model Implementation}
    \label{fig:MetalQuality}
  \end{figure}
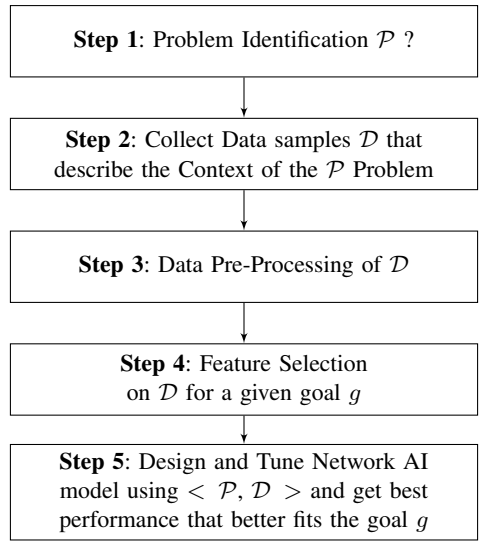

\begin{enumerate}
    \item \textit{Problem Identification:} As mentioned in section~\ref{intro}, the objective of this work consists to analyse the obtained X-Ray diffraction spectra using other programs such as HighScore~\cite{degen2014highscore}, Origin~\cite{may2009software} and to compare them to a specters in the program data base. The spectra properties is resumed in lattice characteristics a, b and grain size D. We notice that the comparison with previous existing results in the literature, enables us to verify the nature of the structure and proves that it concerns tetragonal phase of Tin Oxide SnO2 thin films.
    \item \textit{Data Sample Collection:} To determine a structure for 1d, 2d and 3d materials we must determine stucturals parameters as well as: lattice parameters a, b and c to classified them among the seven systems and fourteen Bravais lattices as mentioned in the crystallography.
    \item \textit{Data Pre-processing:} Data pre-processing techniques are applied, including duplicate detection and removal, missing data imputation, outlier detection and treatment. Data transformation techniques such as normalization and standardization are also employed to ensure data consistency.
    \item \textit{Feature Selection:} Relevant features are selected, with the assistance of domain expert possessing deep knowledge and expertise in the field. These experts provide valuable insights into the key characteristics that define part quality and prioritize the most important features.
    \item \textit{Design and Tuning of Network AI model:} 
    A GRU model is generated after the learning process and its performance are proven using testing.
\end{enumerate}

\subsection{Data Preparation}
\label{dataset}

\begin{itemize}
    \item \textit{Data Description and Collection} \\
    Expert in material physics specify precisely the parameters that should be considered for the characterization of structural materials. Also, all the pipeline tools that is used during the characterization process are identified and configured. After, a data dictionary is constructed as illustrated in the Table~\ref{tab:dataTypes}. Thus, physical symbols and theirs respective designations are specified and validated by material physics experts. We note that this step is crucial and should be done correctly to get an adequate data-driven models that fit precisely the situation of characterization.
    
    \begin{table}[h!]
        \caption{Parameters specification including input or system variables and decision variables, with their respective numerical types. .\\}\label{tab:dataTypes}
        \centering
        \renewcommand{\arraystretch}{1.25}
        \begin{tabular}{l|l|l}
        \hline
        \textbf{Symbol} & \textbf{Definition} & \textbf{Type}\\
        \hline
        \hline
        $\theta$ & Angle between thin films planes and incident beam & float \\
        \hline
        Ndip & Number of dip  & Integer\\ 
        \hline
        Vdip & Velocity of dip & Integer\\ 
        \hline
        Tsec & Drying temperature & Integer\\ 
        \hline
        Trec & Annealing time & Integer\\ 
        \hline
        Time & Drying time & Date\\ 
        \hline
        P$_{sol}$ & Pure SnO2 & Boolean\\ 
        \hline
        D$_{sol}$& Diethanolamine 'DEA' Solution & Boolean\\ 
        \hline
        H$_{sol}$& Hydrochloric acid Solution & Boolean\\ 
        \hline
        A$_{sol}$ & Acetic acid solution & Boolean\\ 
        \hline\hline
        a & Lattice parameter for tetragonal structure in tow direction& float \\
        & x and y & \\
        \hline
        b & Lattice parameter for tetragonal structure in z direction &  float \\
        \hline
        D & Grain size &  float \\
        \hline
    \end{tabular}
    \end{table}

    After, the task of data collection has to be considered with care, knowing its sensitivity and determination in the learning process. It is important to notice here that inaccurate data with poor quality should lead to catastrophic decisions. Many AI models are ineffective because of the lack of date and/or the quality of data.\\

    \item \textit{ Data Pre-processing}\\
    Once the data has been specified and then collected, we can analyse statistically their nature, either by visualization or by computing statistical characteristics, such as the mean $\mu$, the standard variation $\sigma$, the $Median$ and others. This will enable us to choose the best data preparation technique to use - See Table~\ref{tab:dataValues}. We have noticed that the parameters $Tsec(\%s)$ and $Time(\%s)$ are constants because theirs standard deviations are null (See Table~\ref{tab:dataTypes}, $\sigma$= \textbf{0.0}$^\dagger$) and there are multiple scaling in the data set (See Table~\ref{tab:dataTypes}, $\mu(Vdip) = 06.09^* << \mu(Trec)=454.55^*$ and $\sigma(\theta) = 03.68^* << \sigma(Trec)=52.22^*$) and this is not sane to use the data set without normalization and reduction. Thus, we have applied a normalization on the data to have a single scaling for our data and to avoid the multi scaling bias. 
    
    \begin{table}[h!]
        \caption{Description of Data Set distributions of the different features and statistical metrics $\mu$, $\sigma$ and Median.\\}\label{tab:dataValues}
        \centering
        \renewcommand{\arraystretch}{1.25}
        \begin{tabular}{c c | c | c | c | c | c | c | c | c | c || c | c | c}
        \hline
        &$\theta$	&Ndip	&Vdip	&\textbf{Tsec} &Trec &\textbf{Time}	&P$_{sol}$	&D$_{sol}$	&H$_{sol}$	&A$_{sol}$	&a	&b	&D\\
        \hline\hline
        $t_0$&34.2599	&35	&7	&100&500&120	&1	&0	&0	&0	&2.62	&2.61	&99.58\\
        $t_1$&34.36	&7	&7	&100&500&120	&1	&0	&0	&0	&2.61	&2.60	&212.94\\
        $t_2$&27.0841	&35	&7	&100&500&120	&0	&1	&0	&0	&4.65	&2.61	&112.11\\
        $t_3$&27.08	&7	&7	&100&500&120	&0	&1	&0	&0	&4.65	&2.60	&182.07\\
        $t_4$&26.96	&35	&7	&100&500&120	&0	&0	&1	&0	&2.60	&2.60	&602.63\\
        $t_5$&34.4126	&35	&7	&100&500&120	&0	&0	&0	&1	&2.60	&2.60	&602.63\\
        $t_6$&34.0546	&6	&7	&100&400&120	&1	&0	&0	&0	&2.63	&2.63	&1351.28\\
        $t_7$&26.827	&7	&7	&100&400&120	&1	&0	&0	&0	&4.69	&0	&630.41\\
        $t_8$&26.82	&8	&7	&100&400&120	&1	&0	&0	&0	&4.69	&0	&882.48\\
        $t_9$&27.0714	&35	&2	&100&400&120	&1	&0	&0	&0	&4.65	&2.61	&124.65\\
        $t_k$&$\cdots$	&$\cdots$	&$\cdots$	&&$\cdots$&	&$\cdots$	&$\cdots$	&$\cdots$	&$\cdots$	&$\cdots$	&$\cdots$	&$\cdots$\\
        $t_{n}$&27.02	&35	&2	&100&400&120	&0	&1	&0	&0	&4.66	&2.60	&171.09\\
        \hline\hline
        $\mu$ &29.63 &22.27	&\texttt{06.09}$^*$& 100&\texttt{454.55}$^*$&120	&00.55	&00.27	&00.09	&00.09  &04.55	&02.07	&52.22\\
        $\sigma$ &\texttt{03.68}$^*$	&14.63	&02.02	&\textbf{0.0}$^\dagger$&\texttt{52.22}$^*$&\textbf{0.0}$^\dagger$	&00.52	&00.47	&00.30	&00.30	&00.23	&00.69	&51.46\\
        \textit{Median}&27.08	&35.00	&07.00	&100&500.00&120	&01.00	&00.00	&00.00	&00.00	&04.65	&02.28	&31.50\\
        \hline
        \end{tabular}
    \end{table}

    Sometimes and to overcome the lack of the real experimental data, Gaussian augmentation to the dataset are applied. Each data transaction $t$ is used to generate new mutant transactions set \{$t'_k$\} of size $k$ in the neighbor space of some features or attributes in $t$.\\

    \item \textit{ Feature Selection and Reduction}\\
    At this stage of data preparation, features are splitted in input and output variables and some parameters may be just dropped. This techniques called features reduction should be done under the advice of experts in materials physics. In our context, the considered system variables are : $\theta$, $Ndip$, $Vdip$, $Trec$, $P_{sol}$, $D_{sol}$, $H_{sol}$, $A_{sol}$, and the decision variables $a$, $b$, $D$.
    
\end{itemize}

\subsection{GRU Predictor Models: Learning and Testing}
\label{Modeling}
We have generated experimentally the best configuration of the GRU model and the optimal learning parameters are depicted below. In fact, the description of the parameters and their values used along the following experimentation section are introduced right after.   


\begin{itemize}
    \item {\texttt{Tsteps}} (\textit{Time steps refers to the sequence length in the input data.}): \texttt{24}
    \item {\texttt{IN:\#Features + OUT:\#Decisions}} (\textit{Indicates the number of distinct attributes measurements in input.}): \texttt{8 (input parameters) +1 (decision variable features \#$a$, \#$b$, \#$D$).}
    \item {\texttt{Loss}} (\textit{This function measures the dissimilarity between the predicted values and the ground truth values during training.}): \texttt{}
    \item {\texttt{Metrics}} (\textit{These custom functions are used to evaluate the model's performance during training and testing.}): \texttt{MSE, $R^2$, S}.
    \item {\texttt{\#Layer}} (\textit{It denotes the number of layers in the neural network model.}): \texttt{2}
    \item {\texttt{\#Cells}} (\textit{It specifies the number of cells or units per layer of the GRU model.}): \texttt{64}
    \item {\texttt{\#Epochs}} (\textit{This parameter represents the number of times the model iterates over the entire training dataset.}): \texttt{limit <= 800 + StopEarling (SE)}
    \item {\texttt{\#Batch}} (\textit{It refers to the number of samples or data points processed in each training iteration.}): \texttt{}
    \item {\texttt{Act\_func}} (\textit{It is a mathematical function that controls the outputs of each neuron in the GRU neural network.}): \texttt{Relu}
    \item {\texttt{\%Dropout}} (\textit{Dropout is a regularization technique used during training to avoid overfitting and to boost the training.}): \texttt{20\%}
    \item {\texttt{\#FCDense}} (\textit{Known as fully connected layers.}): \texttt{1}
    \item {\texttt{Adam$_{lr}$}}(\textit{They are used to configure the \texttt{lr} learning rate.}): \texttt{lr = 0.0001}
\end{itemize}

\begin{table}
\caption{Tuning hyper-parameters of the learning process {GRU$^a_{64}$}, {GRU$^b_{64}$} and {GRU$^D_{64}$} Models.\\}

\label{table:assembly}
\centering
{\begin{tabular}{l | c | c | c | c | c | c | c | c }
\hline
\textbf{Model} &
\textbf{Metrics} & 
\textbf{Tsteps} &
\textbf{\#Layers} &
\textbf{\#Cells} &
\textbf{\#Epochs (Early Stopping)} &
\textbf{\#Batch} &
\textbf{Act\_func} &
{Adam$_{lr}$}\\ 
\hline
{GRU$^a_{64}$} & MSE/S & 50& 3 & (8, 64, 1) & 450 &  25& Relu &  0.001 \\
\hline
{GRU$^b_{64}$} & MSE/S & 50& 3 & (8, 64, 1) & 480 &  25& Relu &  0.001 \\
\hline
{GRU$^D_{64}$} & MSE/S & 50& 3 & (8, 64, 1) & 270 &  25& Relu &  0.001 \\
\hline
\end{tabular}}
\normalsize
\end{table}

After the tuning of the GRU model training parameters, we will now define an experimental protocol that consists first, in the adoption of the latter fixed learning parameters in Table~\ref{table:assembly}. Second, for each crystal materials characteristics $a$, $b$, and $D$, we have determined respectively the corresponding models as {GRU$^a_{(8,64,1)}$},{GRU$^b_{(8,64,1)}$}, and {GRU$^D_{(8,64,1)}$} - see Table~\ref{table:assembly}. These are trained and then tested using the metrics $MSE$ and $S$ included in the section~\ref{methodos}. 

\begin{table}
\caption{Performance of the GRU models based on Metrics: (a) $MSE$: Mean Square Error, (b) $\mu_{S}$: Mean scores $S$, and (c) Mode$_{S}$: Mode Scores\\}

\label{table:assembly}
\centering
\begin{tabular}{ l | c | c | c | c | c }
\hline
\textbf{Model} &
\textbf{MSE} & \textbf{$\mu_{S}$} & \textbf{Mode$_{S}$} &
\textbf{\#Cells} &
\textbf{Act\_func} \\
\hline
{${GRU}^a_{64}$} & ~7.2901e-14~& ~0.0659~& ~0.5970~&(8, {64}, 1)  & $Relu$ \\
\hline
{${GRU}^b_{64}$} & ~7.7273e-05~& ~0.0675~&~ 0.8797~&(8, {64}, 1)  & $Relu$ \\
\hline
{${GRU}^D_{64}$} & ~1.8649e-09~& ~0.0166~& ~0.1749~&(8, {64}, 1) & $Relu$ \\
\hline
\end{tabular}
\end{table}

\begin{figure}[h!]
        \centering
            \includegraphics[scale=.65]{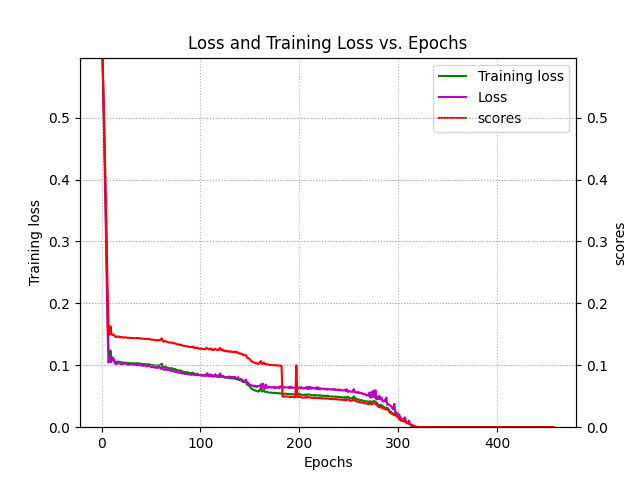}
        \caption{GRU$^{a}_{(8,64,1)}$ Model - Learning loss, Generalization val-loss and Sensitivity scores}
        \label{fig:dim all a properties}
        \end{figure}

\begin{figure}[h!]
            \centering
            \includegraphics[scale=.65]{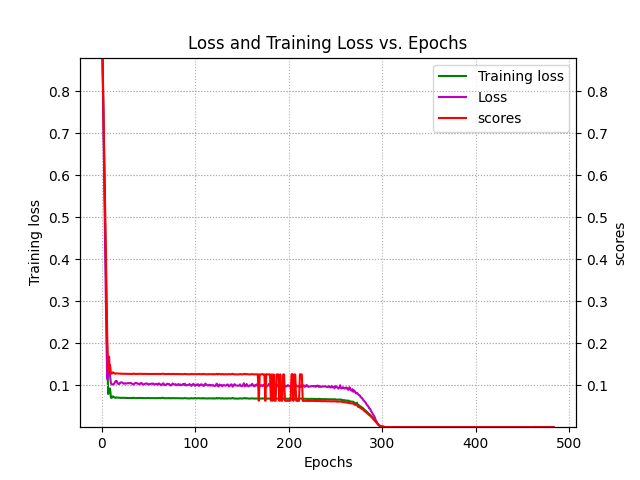}
        \caption{GRU$^{b}_{(8,64,1)}$ Model - Learning loss, Generalization val-loss and Sensitivity scores}
        \label{fig:dim all b properties}
        \end{figure}

\begin{figure}[h!]
            \centering
            \includegraphics[scale=.65]{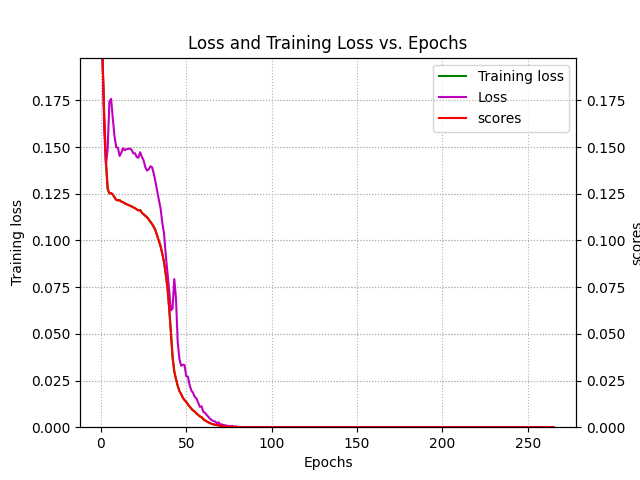}
        \caption{GRU$^{D}_{(8,64,1)}$ Model - Learning loss, Generalization val-loss and Sensitivity scores}
        \label{fig:dim all D properties}
        \end{figure}

At this time, we have depicted the traceability of the training and the testing of our designed models {GRU$^a_{(8,64,1)}$},{GRU$^b_{(8,64,1)}$}, and {GRU$^D_{(8,64,1)}$}. Three figure are presented to define the dynamics of the GRU predictor models. In green curve, we have the learning loss function (\texttt{loss}), and the purple curve represents the generalization or the \texttt{val-loss} function. Finally the red sensitivity function determines whether the learning is stable or not~- see Figures ~\ref{fig:dim all a properties},~\ref{fig:dim all b properties},~\ref{fig:dim all D properties}. 

\begin{itemize}
    \item We notice in the Figure~\ref{fig:dim all a properties}, relative to the generation of the {GRU$^a_{(8,64,1)}$} predictor model for the $a$ crystal geometric feature, that the learning process is unstable until the $200^{em}$ epochs. Before this stage, the learning and generalization are entangled, which means that the learning ans training processes are not yet separated. After that, at the stage of $epochs$ in the range $[20, 310]$, the learning and the generalization seem to be efficient and go to a convergence point around the $310^{em}$ epoch, well with less sensitivity.  Finally, when the $epochs$ parameter is greater than $310$, the dynamics of the training converges to optimal GRU network weight values.\\
    \item In Figure~\ref{fig:dim all b properties}, we can clearly see that the learning in very sensitive in around the $200^{em}$ epoch, but converges around $epoch = 300$ and demonstrates better performances in terms of learning and generalization for the generated model {GRU$^b_{(8,64,1)}$}. \\
    \item For the model {GRU$^D_{(8,64,1)}$} generation - see~Figure~\ref{fig:dim all D properties} -, we note that the learning converges earlier and gives a steady results in epochs $80$ without any sensitivity.\\
\end{itemize}

\subsection{GRU Model Predictions}
\label{ModelPreds}
After the GRU model construction and the performance evaluations for each GRU model predictor of the different structural parameters \#$a$,\#$b$, and \#$D$, we will now apply the predictor models to predict the crystal properties considering by the way the truth experimental values in the dataset dictionary. \\

For each structural features, we use the corresponding model named $GRU^X_{(8,64,1)}$, where $X$ is in the set \{$a$,$b$,$D$\}. Thus, the truth and the estimated properties $a$, $b$, and $D$ are depicted in Table~\ref{tab:Lattice-all} using the above predictor models - see the subsection~\ref{Modeling}. We noticed that in the table~\ref{tab:Lattice-all} (a), the estimated lattice parameter $a$ in the transactions $t_0, t_2$ and $t_6$ are exactly the same as the truth experimental values. The remaining estimated values in \{$t_1$, $t_3$,$t_4$,$t_5$\} have an absolute difference at most equal to $3.00e^{-7}$ of the compared to the truth experimental values. This finding can be observed in the case of the lattices $b$ and $D$. In fact, the largest absolute difference $2.00e^{-2}$ is shown in the transaction $t_5$. \\

\begin{figure}[h!]
    \centering
        \centering
        \includegraphics[scale=0.75]{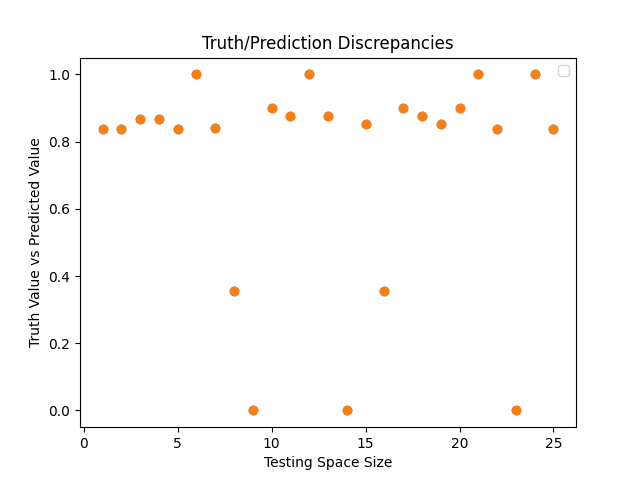}
        \caption{GRU$^{a}_{(8,64,1)}$ Truth-Prediction Discrepancies.}   
        \label{fig:dim a TP}        
\end{figure}

\begin{figure}[h!]
    \centering
        \includegraphics[scale=0.75]{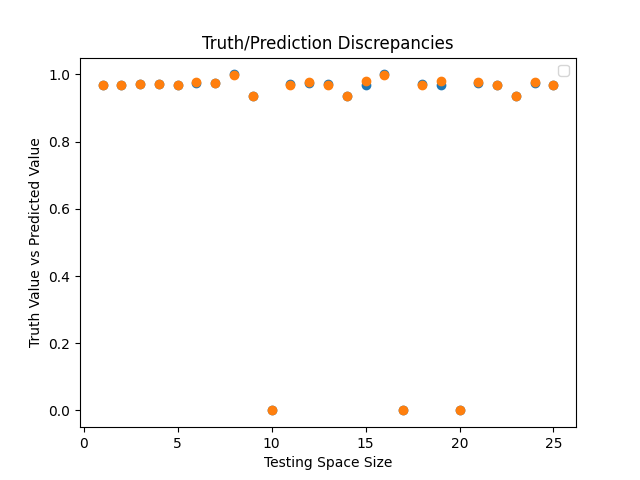}
        \caption{GRU$^{b}_{(8,64,1)}$ Truth-Prediction Discrepancies.}   
        \label{fig:dim b TP}   
\end{figure}

\begin{figure}[h!]
    \centering
        \includegraphics[scale=0.75]{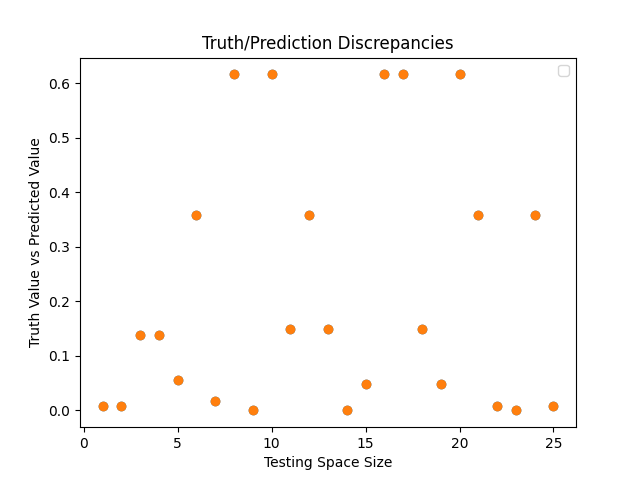}
        \caption{GRU$^{D}_{(8,64,1)}$ Truth-Prediction Discrepancies.}   
        \label{fig:dim D TP}    
\end{figure}

In addition, the difference between the estimated and truth values for the $D$ lattice is in the range of $[4.0e^{-6}, 5.9e^{-5}]$. The Figures~\ref{fig:dim a R2},~\ref{fig:dim b R2},~\ref{fig:dim D R2} below, shows in details the $R^2$ metric which demonstrates the statement that test truth estimated values of $a$, $b$ and $D$ lattices and the predicted computed values are almost equal. In addition, the truth/predicted discrepancies in the Figures~\ref{fig:dim a TP},~\ref{fig:dim b TP},~\ref{fig:dim D TP}, are very small and even zero $0$.    

\begin{figure}[h!]
    \centering
        \includegraphics[scale=0.6]{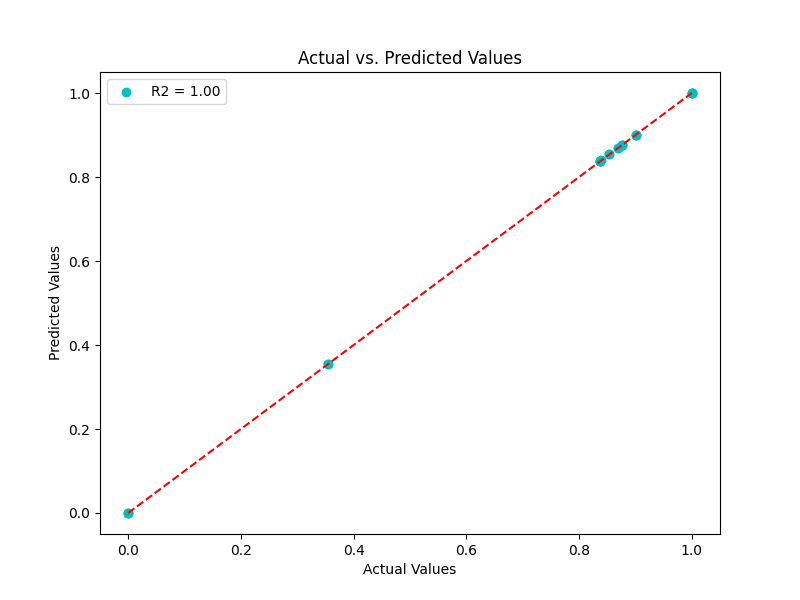}
        \caption{GRU$^{a}_{(8,64,1)}$ Square R2 Metric.}
        \label{fig:dim a R2}
\end{figure}

\begin{figure}[h!]
    \centering
        \includegraphics[scale=0.6]{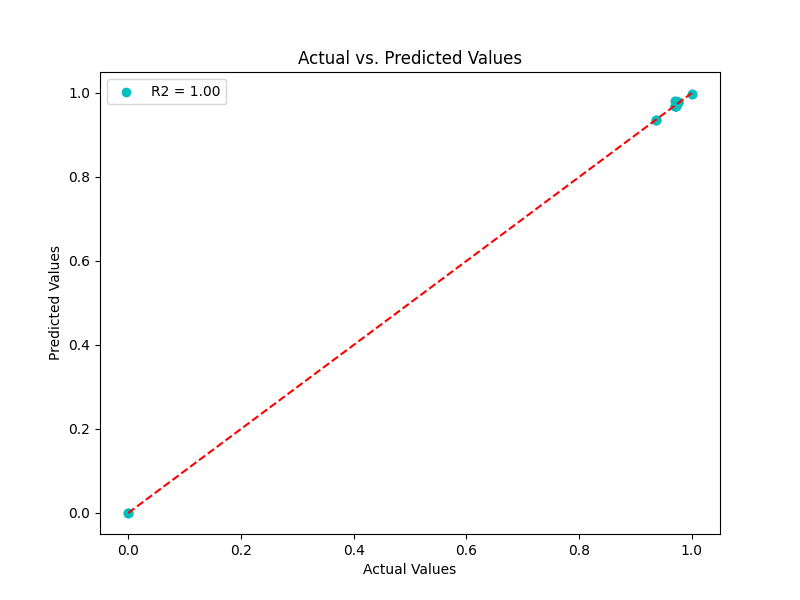}
        \caption{GRU$^{b}_{(8,64,1)}$ Square R2 Metric.}
        \label{fig:dim b R2}
\end{figure}

\begin{figure}[h!]
    \centering
    \centering
        \includegraphics[scale=0.6]{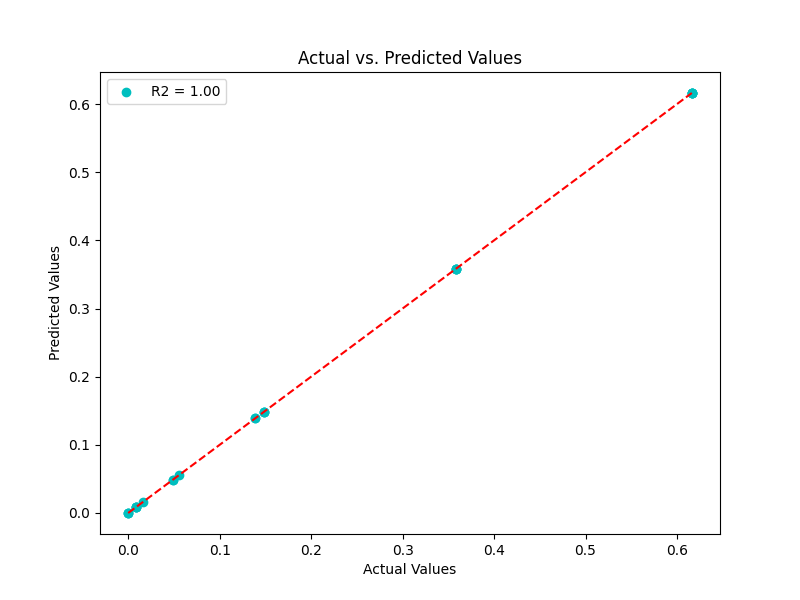}
        \caption{GRU$^{D}_{(8,64,1)}$ Square R2 Metric.}
        \label{fig:dim D R2}
\end{figure}

\begin{sidewaystable}
\begin{subtable}{\textwidth}
        \centering
    \label{tab:Lattice-a}
    \centering
    \small
    \begin{tabular}{ l | l | c | c | c | c | c | c | c || l  | l | c }
    \hline
    $T_n$&{$\theta$} & {$Ndip$}&
    {$Vdip$} & {$Trec$}&
    {$P_{sol}$} & {$D_{sol}$}&
    {$H_{sol}$} & {$A_{sol}$}&

    {Exp$^a$} & {Est$^{a}_{G64}$}&
    $\delta (Exp^a, Est^{a}_{G64})$
    \\
    \hline
    $t_0$&34.3569&7&7&500&1&0&0&0 &4.6509	&~4.6509\textbf{004} &4.00e-07~$\dagger$\\ 
    $t_1$&27.0812&7&7&500&0&1&0&0&4.0830	&~4.0830	&0.00\\
    $t_2$&26.9573&35&7&500&0&0&1&0&4.6719	&~4.6719\textbf{003} &3.00e-07~$\dagger$\\
    $t_3$&34.4126&35&7&500&0&0&0&1&4.7613	&~4.7613	&0.00\\
    $t_4$&34.0546&6&7&400&1&0&0&0&4.6525	&~4.6525 &0.00\\
    $t_5$&26.8223&8&7&400&1&0&0&0&4.6620	&~4.6620	&0.00\\
    $t_6$&27.0714&35&2&400&1&0&0&0&4.6941 &~4.6941\textbf{004}	&4e-07~$\dagger$\\
    \hline
    \hline
\end{tabular}
    \caption{prediction of the Lattice parameter $a$ in $(x,y)$ direction using the GRU $G^a_{64}$ Model. \\}
\end{subtable}
\vspace{0.2cm} 

\centering
\begin{subtable}{\textwidth}
    \label{table:Lattice-b}
    \centering
    \small
    \begin{tabular}{ l | l | c | c | c | c | c | c | c || l  | l | c }
    \hline
    $T_n$&{$\theta$} & {$Ndip$}&
    {$Vdip$} & {$Trec$}&
    {$P_{sol}$} & {$D_{sol}$}&
    {$H_{sol}$} & {$A_{sol}$}&

    {Exp$^b$} & {Est$^{b}_{G64}$}
    &$\delta (Exp^b,Est^{b}_{G64})$
    \\
    \hline
    $t_0$&34.3569&7&7&500&1&0&0&0&2.270\textbf{9}& 	2.270\textbf{6723}&	0.0002277\\
    $t_1$&27.0812&7&7&500&0&1&0&0&2.19\textbf{26}& 	2.19\textbf{11268}&	0.0014732\\
    $t_2$&26.9573&35&7&500&0&0&1&0&2.277\textbf{7}& 	2.277\textbf{169}&	0.0005310\\
    $t_3$&34.4126&35&7&500&0&0&0&1&2.2\textbf{839}& 	2.2\textbf{905123}&	0.0066123\\
    $t_4$&34.0546&6&7&400&1&0&0&0&2.2\textbf{792}& 	2.2\textbf{806101}&	0.0014101\\
    $t_5$&26.827&7&7&400&1&0&0&0&2.2\textbf{703}& 	2.2\textbf{977521}&	0.0274521\\
    $t_6$&27.0714&35&2&400&1&0&0&0&0&	1.1e-04&	0.0001100\\
    \hline
    \hline
    \end{tabular}
    \caption{prediction of the Lattice parameter $b$ in $z$ direction using the GRU $G^b_{64}$ Model. \\
    .}
\end{subtable}
\vspace{0.5cm} 
\begin{subtable}{\textwidth}
    \centering
    \label{table:Lattice-D}
    \centering
    \begin{tabular}{ l | l | c | c | c | c | c | c | c || l  | l | c }
    \hline
    $T_n$&{$\theta$} & {$Ndip$}&
    {$Vdip$} & {$Trec$}&
    {$P_{sol}$} & {$D_{sol}$}&
    {$H_{sol}$} & {$A_{sol}$}&

    {Exp$^D$} & {Est$^{D}_{G64}$}
    &$\delta (Exp^D,Est^{D}_{G64})$
    \\
    \hline
    $t_0$&34.3569&7&7&500&1&0&0&0&18.575&	18.575\textbf{012}	&1.2e-05\\
    $t_1$&27.0812&7&7&500&0&1&0&0&11.3\textbf{8}	&11.3\textbf{79986}	&1.4e-05 \\
    $t_2$&26.9573&35&7&500&0&0&1&0&31.\textbf{5}&	31.\textbf{49995}	&5e-05\\
    $t_3$&34.4126&35&7&500&0&0&0&1&65.32&	65.32\textbf{002}	&2.e-05\\
    $t_4$&34.0546&6&7&400&1&0&0&0&12.633&	12.63\textbf{2996}	&4e-06\\
    $t_5$&26.827&7&7&400&1&0&0&0&17.525&	17.525\textbf{059}	&5.9e-05\\
    $t_6$&27.0714&35&2&400&1&0&0&0&105.1&	105.1\textbf{0001}	&1e-05\\
    \hline
    \end{tabular}
    \caption{prediction of the Gain Size parameter $D$ using the GRU $G^D_{64}$ Model.\\}
\end{subtable}
\caption{{Experimental parameters $Exp^a,Exp^b,Exp^D,$} versus \\
{Estimated $Est^{a}_{G64}$, $Est^{b}_{G64}$, $Est^{D}_{G64}$} values.\\}
\label{tab:Lattice-all}
\end{sidewaystable}

\newpage 

\section{Conclusion}
\label{conclude}
This research work is very beneficial in the sense that it simplifies the hard an complex task of real experiments in materials physics. Indeed, the crucial role of characterization techniques in material physics, particularly emphasizing the significance of gathering data on various physical properties, is not always easy, especially for some methods such as X-ray diffraction that are widely used but still be challenging due to cost, complexity and safety. In brief and to streamline this process, X-ray diffraction outcomes for tin oxide (SnO2) thin films are produced via the Dip-Coating methods deposited onto glass substrates. Then, Various experimental parameters were adjusted to enhance the thin film properties for specific applications. In addition, The structural characteristics, including lattice parameters $a$ and $c$ and the grain size $D$, are computed, confirming the tetragonal phase of thin films in most cases. Finally, due to the extensive array of experimental parameters to be optimized across multiple samples, the use of artificial intelligence can be highly desirable to be considered. We have adopted the use of artificial intelligence techniques, particularly a Gated Recurrent Unit (GRU) model to predict but also estimate structural characteristics and many physical properties of materials.

The relevance and benefits of the proposed data-driven GRU-Based model Predictor for material properties characterization highlight the importance of AI in material physics and leverage discussion about this new pivotal AI methods for understanding and extracting structural properties of crystalline materials. Again, this may show the emphasizing on AI and its role in characterizing material properties across various fields, when limitations in terms of computation capacity for ab-initio calculation ans simulation methods. Our GRU models have gained prominence in enhancing X-ray diffraction's capabilities and strengthening predictions of materials' structural properties in an easy and safe thin films SnO2 characterization. Furthermore, we proposed a data-driven framework for material characterization, focusing on the proposed AI GRU model among others. In the soon future, we will continue exploit derive crystal properties through experimental data to design specific AI models that can facilitate and revolutionize the field of material science and characterization, offering solutions to the challenges faced by engineers and researchers when using traditional techniques and computational constraints.

\bibliographystyle{splncs04}
\bibliography{references.bib}    

\end{document}